\documentclass[iop]{emulateapj}
\usepackage{natbib} 
\def\lapp{\ifmmode\stackrel{<}{_{\sim}}\else$\stackrel{<}{_{\sim}}$\fi}
\def\gapp{\ifmmode\stackrel{>}{_{\sim}}\else$\stackrel{>}{_{\sim}}$\fi}
\usepackage{multirow}
\usepackage{color}
\usepackage{subfigure}
\usepackage{bm}
\usepackage{lmodern}
\usepackage{soul}
\usepackage{amsmath}

\begin{document}

\title{NuSTAR Observations of the Young, Energetic Radio Pulsar PSR B1509$-$58}

\author{Ge Chen\altaffilmark{1}, Hongjun An\altaffilmark{2}, Victoria M. Kaspi\altaffilmark{1, 5}, Fiona A. Harrison\altaffilmark{3}, Kristin K. Madsen\altaffilmark{3}, Daniel Stern\altaffilmark{4}
}
\affil{
{\small $^1$Department of Physics, McGill University, Montreal, Quebec, H3A 2T8, Canada}\\
{\small $^2$ Department of Physics/KIPAC, Stanford University, Stanford, CA 94305-4060, USA}\\
{\small $^3$ Cahill Center for Astronomy and Astrophysics, California Institute of Technology, Pasadena, CA 91125, USA}\\
{\small $^4$ Jet Propulsion Laboratory, California Institute of Technology, 4800 Oak Grove Drive, Mail Stop 169-221, Pasadena, CA 91109, USA}\\
}
\altaffiltext{5}{Lorne Trottier Chair; Canada Research Chair}
\begin{abstract}
We report on {\em Nuclear Spectroscopic Telescope Array (NuSTAR)} hard X-ray observations of the young rotation-powered radio pulsar PSR B1509$-$59 in the supernova remnant MSH 15$-$5{\it 2}.  We confirm the previously reported curvature in the hard X-ray spectrum, showing that a log parabolic model provides a statistically superior fit to the spectrum compared with the standard power law. The log parabolic model describes the {\it NuSTAR} data, as well as previously published $\gamma$-ray data obtained with COMPTEL and {\it AGILE}, all together spanning 3 keV through 500 MeV. Our spectral modelling allows us to constrain the peak of the broadband high energy spectrum to be at 2.6$\pm$0.8~MeV, an improvement of nearly an order of magnitude in precision over previous measurements. In addition, we calculate {\em NuSTAR} spectra in 26 pulse phase bins and confirm previously reported variations of photon indices with phase. Finally, we measure the pulsed fraction of PSR B1509$-$58 in the hard X-ray energy band for the first time. Using the energy resolved pulsed fraction results, we estimate that the pulsar's off-pulse emission has a photon index value between 1.26 and 1.96. Our results support a model in which the pulsar's lack of GeV emission is due to viewing geometry, with the X-rays originating from synchrotron emission from secondary pairs in the magnetosphere.

\end{abstract}

\keywords{pulsars: individual (PSR B1509$-$58) -- stars: neutron -- X-rays: stars}

\section{Introduction}
\label{sec:intro}

PSR B1509$-$58 is a young, energetic rotation-powered pulsar discovered in the soft X-ray band using the {\em Einstein Observatory} (\citealt{1982ApJ...256L..45S}) and soon afterwards detected in the radio band (\citealt{1982ApJ...262L..31M}). It has period $\sim$150 ms, period derivative  $\sim$1.5 $\times10^{-12} \rm s \ s^{-1}$ and a high spin-down luminosity of $\dot E=1.7\times10^{37} \rm erg~s^{-1}$. Assuming the standard magnetic dipole model, its characteristic spin-down age is $\sim$1600 years and inferred surface magnetic field at the equator is  $\sim1.5\times10^{13}$ G, higher than the lowest estimation of $B_{\rm dip}\simeq6\times10^{12}\rm G$ for a magnetar (\citealt{2013ApJ...770...65R}). The pulsar lies in the centre of the supernova remnant MSH 15$-$5{\it 2}, which has a complex structure with thermal and non-thermal emission, which is largely powered by the pulsar (\citealt{1996PASJ...48L..33T}). Recently, {\it NuSTAR} observations of the MSH 15$-$5{\it 2} and its pulsar wind nebula (PWN) region by \cite{an2014high} revealed clear evidence for synchrotron burnoff in the PWN, spectral softening with distance from the pulsar, and an interesting shell-like structure in the $N_{\rm H}$ map.

The aforementioned {\it NuSTAR} observations also offer a new window into the hard X-ray emission of the central pulsar itself, the subject of this work. PSR B1509$-$58 is one of the brightest rotation-powered X-ray pulsars in the sky and has been observed by many X-ray and soft $\gamma$-ray telescopes, for example, {\em Ginga} (\citealt{1993AIPC..280..213K}), {\em OSSE} (\citealt{1994ApJ...434..288M}), {\em WELCOME} (\citealt{1994ApJ...428..284G}), {\em RXTE} (\citealt{marsden1997x}), COMPTEL (\citealt{1999A&A...351..119K}), {\em AGILE} (\citealt{2010ApJ...723..707P}), {\em BeppoSAX} (\citealt{2001A&A...375..397C}), and \textit{Fermi} LAT (den Hartog et al. in preparation). Thanks to {\em NuSTAR}'s wide energy range in the hard X-ray band and unique hard X-ray focusing ability therein, {\em NuSTAR} observations of PSR B1509$-$58 can shed new light on the pulsar's hard X-ray spectrum.

The pulsar's spectrum in the hard X-ray band has long been of interest, in part because the source's large flux enables detailed studies not possible in most rotation-powered pulsars. Based on {\em BeppoSAX} and COMPTEL observations, \cite{2001A&A...375..397C} reported that the spectral energy distribution of PSR B1509$-$58 is well represented by a logarithmic parabolic function characterized by a linear dependence of the spectral slope upon the logarithm of energy from $\sim$1 keV to $\sim$30 MeV, rather than the simple power-law model so ubiquitously assumed for rotation-powered pulsar spectra. The spectral bending in PSR B1509$-$58 was reported to be detectable also in the keV energy range. If correct, this has important implications for the high energy emission mechanism. However, {\em BeppoSAX} was an X-ray astronomy satellite that consisted of three instruments operating in different energy bands (\citealt{1997A&AS..122..299B}), and slight systematic cross-calibration discrepancies among these three instruments have been reported by \citet{2005SPIE.5898...22K}. Hence confirmation of the claim of spectral curvature from {\it BeppoSAX} data is important to obtain, particularly with a mission like {\it NuSTAR} which covers the full 3--79~keV band with a single instrument. This, when combined with recent soft $\gamma$-ray observations, can in principle provide vastly improved constraints on the full high-energy spectrum, which could help better determine the peak spectral energy and ultimately address the curious low-energy cutoff seen in the high-energy $\gamma$-ray spectrum from this source (\citealt{2010ApJ...714..927A}).

In this paper, we present timing and spectral analyses of PSR B1509$-$58 data obtained by {\em NuSTAR} as well as the {\it Chandra X-ray Observatory}. In Section \ref{sec:obs}, we describe our observations and data reduction. In Section \ref{sec:ana}, we present our analysis and results. In Section \ref{sec:disc} we discuss the results and in Section \ref{sec:sum} provide a summary.

\section{Observations}
\label{sec:obs}
{\em NuSTAR} is the first focusing X-ray telescope that operates above 10 keV (\citealt{2013ApJ...770..103H}). It consists of two co-aligned hard X-ray grazing incidence telescopes, which focus onto two independent solid-state focal plan modules, and produce two spectra in each observation. The angular resolution of the observatory is 18$''$ FWHM, with a half power diameter of 58$''$. Both telescopes operate in the energy band from 3 to 79 keV and each provides an energy resolution of 400 eV FWHM at 10 keV and 0.9 keV at 60 keV. The temporal resolution is 2 $\mu$s, and the timing accuracy is 1--2 ms.

Four {\em NuSTAR} observations were made of the region containing PSR B1509$-$58 as well as its PWN in June and August of 2013, with a total exposure time of 178.9 ks (see \citealt{an2014high}). In every observation, the two telescopes each produced a spectrum for a total of eight spectra. 
We also analyzed an archival {\em Chandra} observation made with the High Resolution Camera (HRC) in June 2005, with an exposure time of 44.9 ks. 
The HRC is sensitive within 0.08--10 keV. Since the spectral response of HRC is extremely limited, we used the {\em Chandra} data only for the timing analysis below.
Table \ref{ta:obs} summarizes the observations.

\begin{table*}[]
\begin{center}
\caption{Summary of Observations}
\label{ta:obs}
\begin{tabular}{cccccc}
\hline\hline
Observatory  &  Obs. ID  &  Start Epoch&  Start Date    & Stop Date &  Exposure   \\
		   & 		      &	    (MJD)  	    &	(UT)   	    &	(UT)		& (ks)  	    \\
\hline
{\em NuSTAR} & 40024004002 & 56450.9 & 2013Jun07 & 2013Jun08& 43.4  \\
{\em NuSTAR} & 40024002001 & 56451.8 &2013Jun08 & 2013Jun09& 42.6  \\
{\em NuSTAR} & 40024003001 & 56452.6 & 2013Jun09 & 2013Jun10& 44.3  \\
{\em NuSTAR} & 40024001002 & 56519.6 & 2013Aug15 & 2013Aug16  & 48.6  \\
{\em Chandra} (HRC) & 5515 & 53534.0 & 2005Jun13 & 2005Jun13 & 44.9  \\
\hline\hline
\end{tabular}
\end{center}
\end{table*}

The {\em NuSTAR} data were processed with {\ttfamily HEASOFT} v6.15.1, {\ttfamily nupipeline} v1.3.1, and the Calibration Database (CALDB) files from 20131223. Standard Level 2 cleaned events files were generated for further analysis. 
The {\em Chandra} data were reprocessed with {\em chandra\_repro} in CIAO 4.5 in order to use the most recent calibration files.

\section{Data Analysis and Results}
\label{sec:ana}

\subsection{Timing Analysis}
\label{sec:timingana}
From each {\em NuSTAR} observation, pulsar events were extracted in the 3--79 keV energy band using a circular region of radius 30$''$ centered on the pulsar. 
We then applied a correction to convert photon arrival times to the equivalent time at the solar system barycenter (Barycentric Dynamical Time) using the DE200 ephemeris.
We verified that the pulse periods at each epoch were consistent with those predicted by the ephemeris of \citet{2011ApJ...742...31L}.

To measure the properties of the X-ray pulsations of PSR B1509$-$58, the background must be carefully subtracted. The background consists of two components: detector background and PWN. 
For the {\em NuSTAR} observations, the detector background was extracted with an aperture of radius 45$''$ in a pulsar-free and PWN-free region on the same detector chip with the pulsar. We also tried different background regions and found that all results in our analysis were independent of the exact choice of region. 
The PWN, however, is important to consider because of {\em NuSTAR}'s broad point spread function (PSF), which will be described later in this section.

\begin{figure}[]
  \centering
  \includegraphics[width=0.5\textwidth]{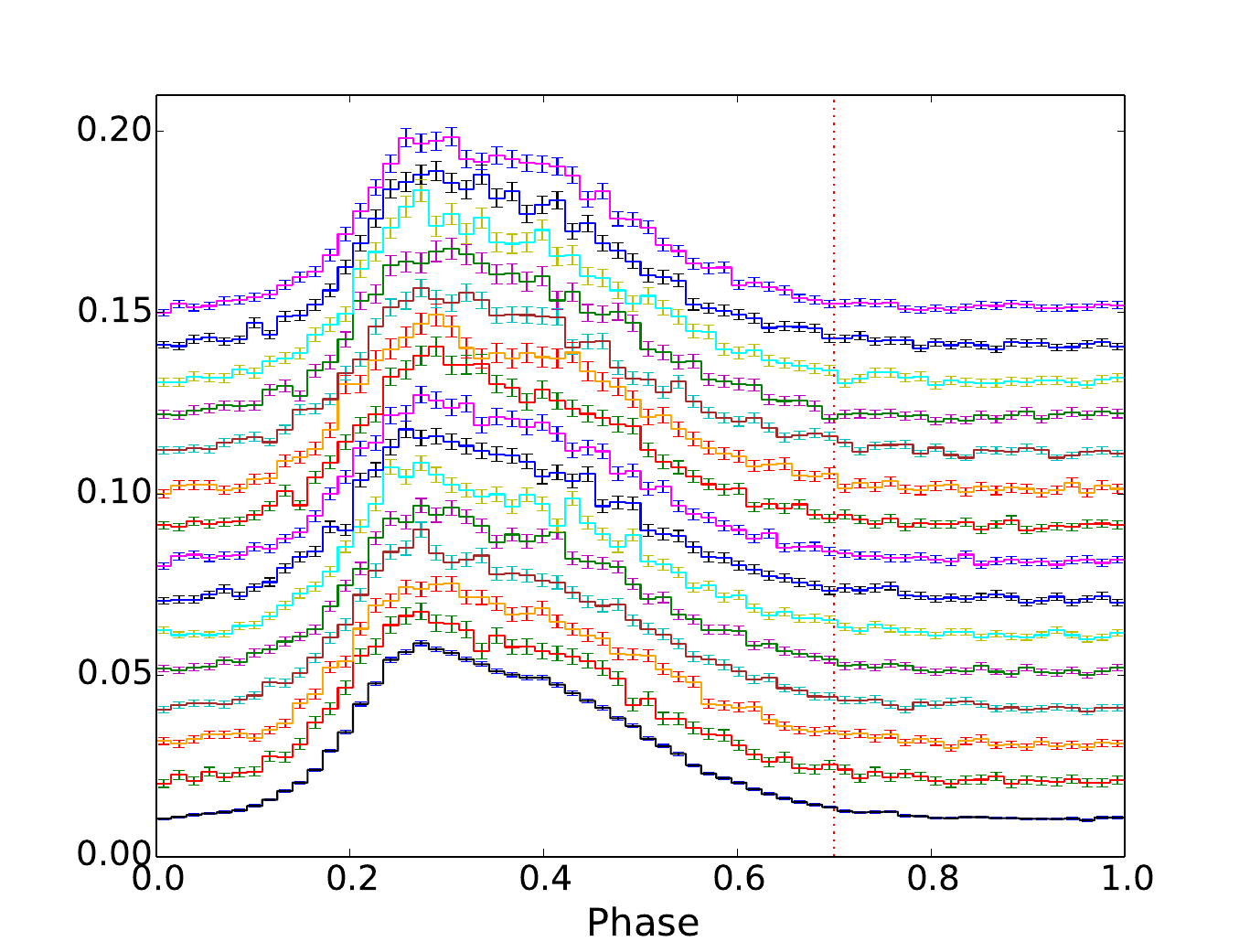}
  \caption{Pulse profiles from {\em NuSTAR} observations in various energy bands: from bottom to top are energy band of 3--79 keV, 3--4 keV, 4--5 keV, 5--6 keV, 6--7 keV, 7--8 keV, 8--9 keV, 9--10 keV, 10--11 keV, 11--12 keV, 12--14 keV, 14--16 keV, 16--19 keV, 19--25 keV and 25--79 keV, respectively. Phase zero was selected approximately around the point where the pulsation starts. The histograms each contain 64 bins. Values of the y-axis don't have physical meaning. No significant change in shape was found in different energy bandpasses. We used phase 0.7--1 (vertical dotted line) as the off-pulse component in Section \ref{sec:spectrumana}. 
\label{fig:pp}
}
\end{figure}

We produced pulse profiles using the {\em NuSTAR} data in 14 energy bandpasses from 3--79 keV (Fig. \ref{fig:pp}). Each pulse profile contained 64 bins and the number of events in each bin was at least 20. To obtain the pulsed component from a pulse profile, we determined the smallest photon count value among the 64 bins, used it as the off-pulse level, and subtracted this off-pulse value from each of the 64 bins.

In order to search for pulse profile evolution with energy, we compared the shapes of the pulse profiles in the 14 energy bands. We scaled them by dividing each with the total photon counts in that pulse profile. We chose the pulse profile of 3--79 keV as a reference, subtracted each of the 14 normalized profiles from it, and tested $\chi^2$. No significant variation in shape was detected, and the residuals appeared to be random. The $\chi^2$ test showed that the probability of the pulse profiles being the same in shape with the reference is above 40\% for 11 energy bands. The most suspicious one had a probability of 0.24\%, still not sufficiently small to indicate a disagreement in shape, considering the number of trials. Furthermore, shifts in phase could indicate that emission regions vary with energy. To check this, we performed a cross correlation of each pulse profile with the reference, but no significant phase shift was seen for any profile.

The pulsed fraction can be defined in several ways. Here we chose the RMS pulsed fraction, as well as an intuitive definition based on the area under the pulse. The RMS pulsed fraction is defined as (\citealt{dib2009rossi})
$$PF_{\rm RMS}=\frac{\sqrt{2\sum_{k=1}^{8}((a_k^2+b_k^2)-(\sigma_{a_k}^2+\sigma_{b_k}^2))}}{a_0}.$$
Here, $a_k$=$\frac{1}{N}\sum_{i=1}^{N}r_i \cos(2\pi \phi_i)$, $b_k$=$\frac{1}{N}\sum_{i=1}^{N}r_i \sin(2\pi \phi_i)$, and $\sigma_{a_k}$, $\sigma_{b_k}$ are the uncertainties in $a_k$ and $b_k$ respectively. ${\phi_i}$ is the phase, $N$ is the total number of phase bins, and $k$ refers to the number of Fourier harmonics. 

The area pulsed fraction is 
$$PF_{\rm area} = \frac{\sum_{i=1}^{N}(r_i-r_{\rm min})}{\sum_{i=1}^{N}r_i}.$$
Here, ${r_i}$ is the count number in the $i$th phase bin, $N$ is the total number of phase bins, and $r_{min}$ is the minimal count number in all bins.

\begin{figure}[]
\includegraphics[width=0.5\textwidth]{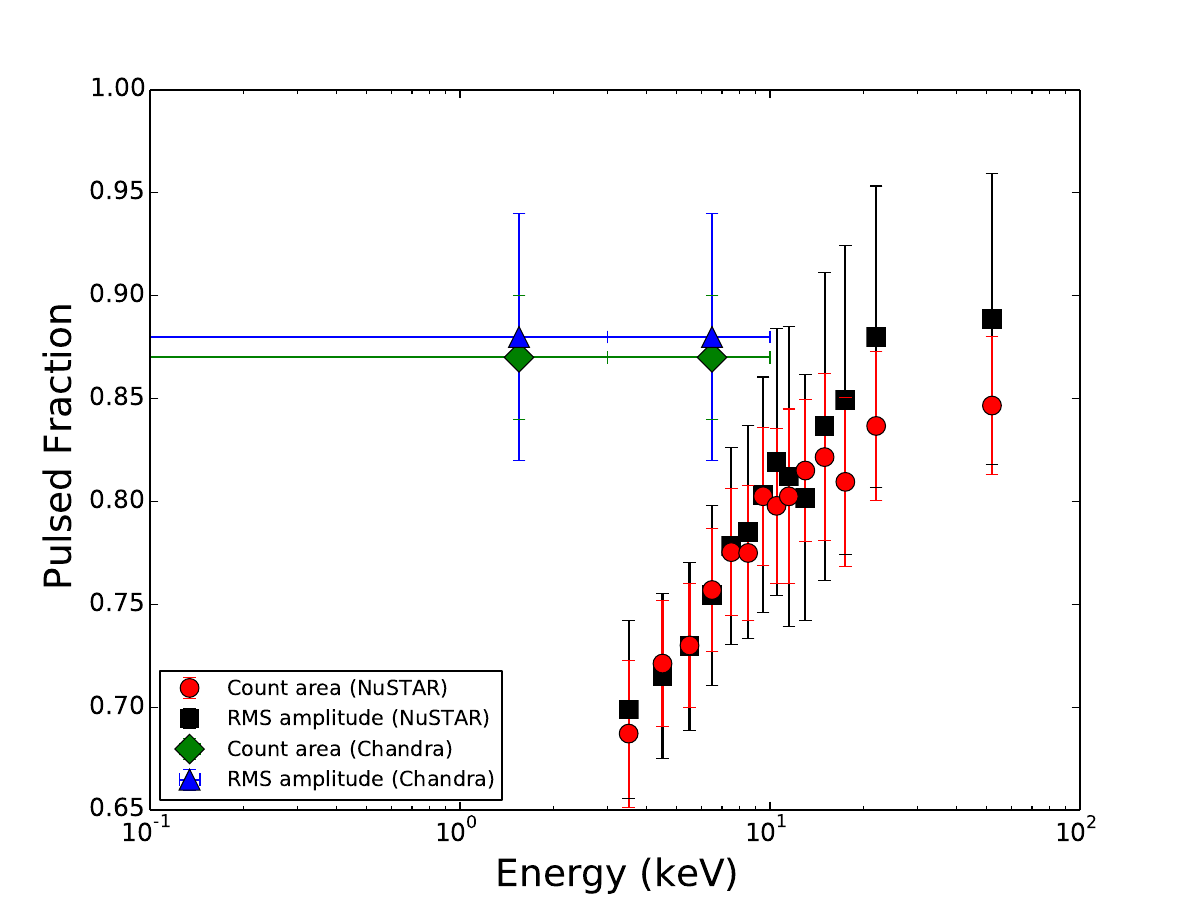}
\caption{RMS and area pulsed fractions in multiple energy bands from {\em NuSTAR} and {\em Chandra} observations. On the one hand, the {\em NuSTAR} results increase significantly between 3 to $\sim$19 keV and then approach a plateau within 19--79 keV: the RMS pulsed fraction of the highest energy band is about 3$\sigma$ larger than that of the lowest one. On the other hand, the {\em Chandra} results do not show any variation between 0.08--10 keV. Moreover, the RMS pulsed fraction of {\em Chandra} is consistent with {\em NuSTAR}'s results above 19 keV within 0.2$\sigma$.
\label{fig:PF}
}
\end{figure}

Figure \ref{fig:PF} displays the RMS and area pulsed fractions measured for pulse profiles in the 14 energy bands. Both the RMS and area pulsed fractions increase with energy between 3 and $\sim$19 keV and then approach a plateau at 19--79 keV. The RMS of the highest energy band is about 3$\sigma$ larger than that of the lowest band. 

To estimate the pulsed fraction that is free from PWN contamination (PWN-free), pulsed fractions were also measured from the {\em Chandra} HRC observation. The RMF of the {\em Chandra} HRC is complicated, so the energy-resolved HRC pulsed fractions (Fig. \ref{fig:PF}) should be regarded as 
as having been measured just in ``soft'' and ``hard'' bands, not the exact energy ranges. This does not change any result though, since we found the {\em Chandra} pulsed fraction value to be constant in these two bands.
For the {\it Chandra} observation, we can ignore background contamination because of {\it Chandra}'s narrow PSF. Specifically, in the {\it Chandra} observation within 0.08--10 keV, there were 12100 events collected in the pulsar region ($R$=1.5$''$) while the number of background events collected within an annular region with $R_{in}$=2$''$ and $R_{out}$=4$''$ was 374. Hence the background contamination in the {\it Chandra} pulsar region would be 70/12100, which is ignorable, and the PWN-free pulsed fraction value could be measured directly. 

We estimated the PWN contamination in 3--10 keV in the {\em NuSTAR} pulsar region using the {\em Chandra} HRC observation. In 0.08--10 keV, the {\em Chandra} observation had 8080 counts in an annular region with $R_{in}$=1.5$''$ and $R_{out}$=30$''$. We used PIMMS\footnote{http://heasarc.gsfc.nasa.gov/docs/software/tools/pimms.html} to convert this into the 3--10 keV {\em NuSTAR} counts assuming a power-law spectrum with a photon index of 1.7 (\citealt{an2014high}), and the result was 12694 counts. In 3--10 keV, the {\em NuSTAR} data had 37927 counts in the pulsar region, so the PWN contamination is roughly 12694/37927, not ignorable. This implied a pulsed fraction of $\sim$66\%, consistent within 3$\sigma$ with our {\em NuSTAR} measurements in 3--4 keV, 4--5 keV, 5--6 keV, 6--7 keV, 7--8 keV, 8--9 keV and 9--10 keV.



Figure \ref{fig:PF} shows that the RMS and area pulsed fractions measured from {\em Chandra} HRC between 0.5--10 keV are higher than the results measured from {\em NuSTAR} in 3--10 keV, however, the {\em Chandra} results are consistent with the {\em NuSTAR} results above 19 keV to within 0.2$\sigma$. This suggests that the variation of pulsed fraction from {\em NuSTAR} is due to the variation in the PWN contribution, which decreases faster than the pulsar spectrum at higher energy. In Section \ref{sec:disc}, we will use these results to estimate contributions from the PWN and the pulsar's off-pulse component, and discuss constraints on the photon index of the pulsar's off-pulse component.
\\
\\
\subsection{Phase-averaged and Pulsed Spectral Analysis}

\label{sec:spectrumana}
\begin{table}[]
\begin{center}
\caption{Cross Normalization Factors of Each Spectrum}
\label{ta:cross_norm}
\begin{tabular}{cc}
\hline\hline
Obs. ID & Cross Normalization Factor \\
\hline
40024004002A & 1 \\
40024004002B & $1.00\pm0.014$ \\
40024002001A & $0.789^{+0.014}_{-0.013}$ \\
40024002001B & $0.794\pm0.013$ \\
40024003001A & $1.02\pm0.015$ \\
40024003001B & $1.04\pm0.015$ \\
40024001002A & $1.02\pm0.016$ \\
40024001002B & $1.03\pm0.016$ \\
 \hline\hline
\end{tabular}
\end{center}
\end{table}

\begin{table*}
\begin{center}
\caption{Pulsed and Phase-averaged Spectral Analysis Results}
\label{ta:pulsed}
\begin{tabular}{cccccccc}
\hline\hline
Model  &Photon Index & {$\alpha$} & {$\beta$} & $K$ & Flux (3-79 keV) & $\chi^2/dof$ & Null Probability \\
 & & & & [$10^{-3} \rm ph~cm^{-2}~s^{-1} keV ^{-1}$] & [$10^{-10} \rm erg~cm^{-2}~s^{-1}$] & & \\
\hline
Power Law & $1.386\pm0.007$ &... &... &  $3.70\pm0.07$ & $1.198\pm0.014$  & 278/254 &14.4\% \\
\hline
Logpar & ...& $1.16\pm 0.05$ & $0.11\pm0.02$ & $2.81^{+0.16}_{-0.15}$ & $1.136\pm0.018$ & 254/253 & $47.6\%$ \\
\hline
Logpar\footnote{These parameters are from Cusunamo et al. (2001) in the 2-- 300 keV band.} & ...& $0.96\pm0.08$ & $0.16\pm0.04$ & $1.8\pm0.3$ &$\sim1.33 $   &...&...\\
\hline
\end{tabular}
\end{center}
\end{table*}

\begin{table*}
\begin{center}
\caption{Broadband Spectral Analysis Results at Different Energy Bands}
\label{ta:broadband}
\begin{tabular}{ccccccc}
\hline\hline
Energy & $\alpha^a$ & $\beta^a$ & $K^a$ &  $\chi^2/dof$ & Null Probability  \\
 & & & [$10^{-3} \rm ph~cm^{-2}~s^{-1} keV ^{-1}$] & & \\
\hline
{\em NuSTAR} Results &&&&&\\
3--79 keV & $1.09\pm 0.04$ & $0.13\pm0.02$ & $2.47\pm0.12$ & 295/260 & $7\%$ \\
3 keV--30 MeV & $1.104\pm 0.018$ & $0.124\pm0.007$ & $2.51\pm0.06$ & 298/263 & $7\%$ \\
3 keV--500 MeV & $1.083\pm 0.015$ & $0.134\pm0.005$ & $2.45\pm0.06$ & 301/265 & $6\%$ \\
\hline
Cusumano et al. (2001) &&&&&\\
2-300 keV& $0.96\pm0.08$ & $0.16\pm0.04$ & $1.8\pm0.3$  &...&...\\
2 keV--30 MeV& $1.03\pm0.05$ & $0.13\pm0.02$ & $1.9\pm0.1$  &...&...\\
\hline
\footnotetext{The parameters $\alpha$, $\beta$ and $K$ were defined in Equation \ref{eq:logpar}.}
\end{tabular}
\end{center}
\end{table*}

In order to examine the spectrum of the pulsed emission from the pulsar, we first defined phase interval 0--0.7 as the pulsed component and the mean count rate between phases 0.7--1.0 as the off-pulse component (see Fig.\ref{fig:pp}). We extracted 3--79 keV pulsar events and produced pulsed spectra. The off-pulse level was then subtracted from the pulsed component to yield the pulsed spectrum. 
In every observation, the two telescopes each produced a spectrum, so we had eight spectra from these four {\em NuSTAR} observations (Section \ref{sec:obs}). Each of the eight spectra was grouped into at least 600 counts per energy bin and then fitted jointly to various models, using the software package {\em XSPEC} v12.8.1g and using $\chi^2$ minimization. To check the results, we also grouped the spectra into 20, 50, 100, 200 and 400 counts per bin. In each case, we used both the $\chi^2$ minimization, and the {\ttfamily cstat} statistic, which is for Poisson data. The results did not differ significantly, so we only report the results for the spectra that were grouped into 600 counts per bin, and fitted with $\chi^2$ minimization statistics. 

\begin{figure}[]
  \centering
  \subfigure[]{\label{fig:pow}\includegraphics[width=0.5\textwidth]{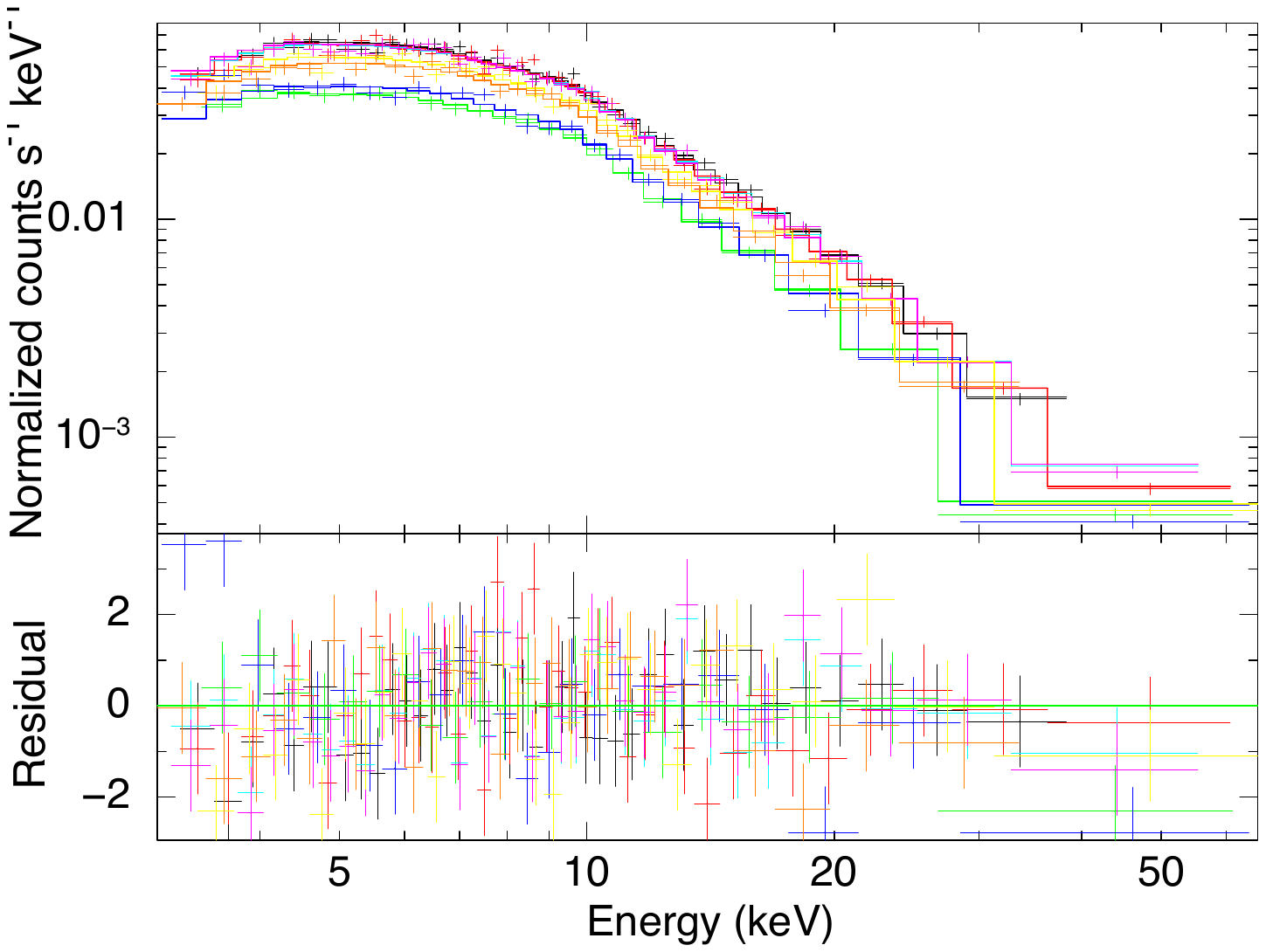}}
  \subfigure[]{\label{fig:logpar}\includegraphics[width=0.5\textwidth]{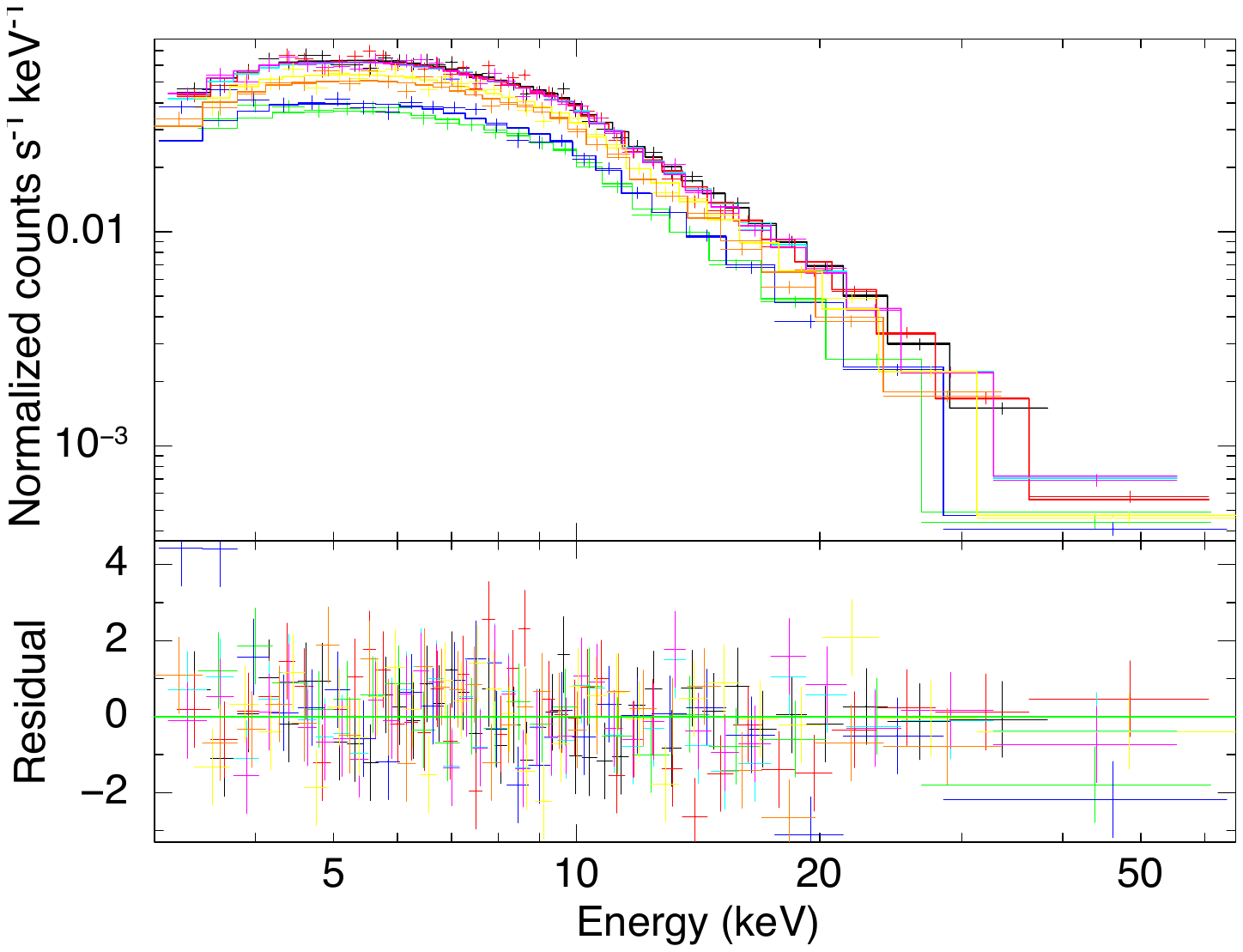}}
  \caption{(a) Top panel: {\em NuSTAR} observation data and the power-law model. Bottom panel: residual (data $-$ model). The power-law model is consistent with our data, $\chi^2$ per dof is 277.989/254 ($p$ = 14.41\%). Note that a systematically curved trend is seen in the residuals. (b) Top panel: {\em NuSTAR} observation data and the logpar model. Bottom panel: residual (data $-$ model). The logpar model fits with our data better. $\chi^2$ per dof is 253.662/253 ($p$ = 47.64\%). No visible trend is seen in the residuals. Eight Colors correspond to the eight spectra used in this analysis.
\label{fig:models}
}
\end{figure}

We first fitted the pulsed spectra with an absorbed single power-law model over 3--79 keV. 
Since the {\em NuSTAR} data are not sensitive to a small change in the equivalent hydrogen column $N_{\rm H}$, we fixed it at the previously reported value $0.95\times 10^{22}\rm \ cm^{-2}$ (Gaensler et al. 2002). Cross normalization factors were introduced for each spectrum because each observation sampled the source in a different detector region, as shown in Table \ref{ta:cross_norm}. Two of the normalization factors were below 1, since part of the pulsar field fell into the detectors' chip gaps. We then tied all the fitting parameters except the cross normalization factors for each spectrum. Table \ref{ta:pulsed} shows the result. Figure \ref{fig:pow} shows the spectra and the fitted single power-law model. Since we used phase 0--0.7 to produce the pulsed spectra, the exposure time was 0.7 times smaller than it should be. We corrected the normalization parameter $K$ and flux results throughout the spectral analysis with a factor of $1/0.7$. With $\chi^2$ per degree of freedom (dof) 278/254, the null hypothesis that the power-law model describes our data had a probability of $p$ = 14.41\%, so the result was acceptable. However, note in Figure \ref{fig:pow} that a systematically curved trend is seen in the residuals. Curved trends are also visible in the residuals of the spectra grouped into at least 400 counts per bin, and of the rebinned plots of the spectra grouped into at least 200, 100, 50 and 20 counts per bin. These imply that the single power-law model may not be the optimal representation of our data. 

Following the work of \citet{2001A&A...375..397C}, we next fitted the spectra with a logarithmic parabolic (logpar) model: 
\begin{equation}
\label{eq:logpar} 
A(E) = K(E/E_{\rm p})^{-\alpha-\beta log(E/E_{\rm p}))}.
\end{equation}
Here, $\alpha$ is the slope at the pivot energy ($E_{\rm p}$), $\beta$ is the curvature term, $K$ is the normalization factor and we fixed $E_{\rm p}$ at 1 keV, as \citet{2001A&A...375..397C} did. The values obtained were $\alpha=1.16\pm0.05$, $\beta=0.11\pm0.01$ and $K=(2.81^{+0.16}_{-0.15})\times10^{-3}$ ph cm$^{-2}$ s$^{-1}$ keV$^{-1}$. A positive $\beta$ value means that the spectrum becomes softer at higher energy.
Table \ref{ta:pulsed} and Figure \ref{fig:logpar} show the results and spectra. With $\chi^2$ per dof 254/253, the null hypothesis that the logpar model describes our data had a probability of $p$= 47.64\%, and the model was acceptable.  

To determine whether the improvement in $\chi^2$ using the logpar model is statistically valid, we performed the standard F-test to estimate the probability that the improvement in $\chi^2$ tests was by chance. We found the probability to be $1.5\times10^{-6}$, so the logpar model is statistically a better description of the data than the power law. Also the residuals in Figure \ref{fig:logpar} do not show an obvious trend. To test the probability that the residuals were random, we performed the Runs test (also known as Wald--Wolfowitz test; \citealt{2012ApJ...748...86O}; \citealt{Barlow:0471922951}) in {\em XSPEC}. The result was 2\% for the power-law model, so the curved trend seen in the residuals of Figure \ref{fig:pow} is statistically significant. For the logpar model, the result was 8\%, so statistically the residuals of Figure \ref{fig:logpar} are 4 times more likely to be randomly distributed than those of Figure \ref{fig:pow}. Hence, the logpar model provides a statistically superior fit to the spectrum compared with the standard power law. 

\begin{figure}[]
\centering
\includegraphics[width=0.5\textwidth]{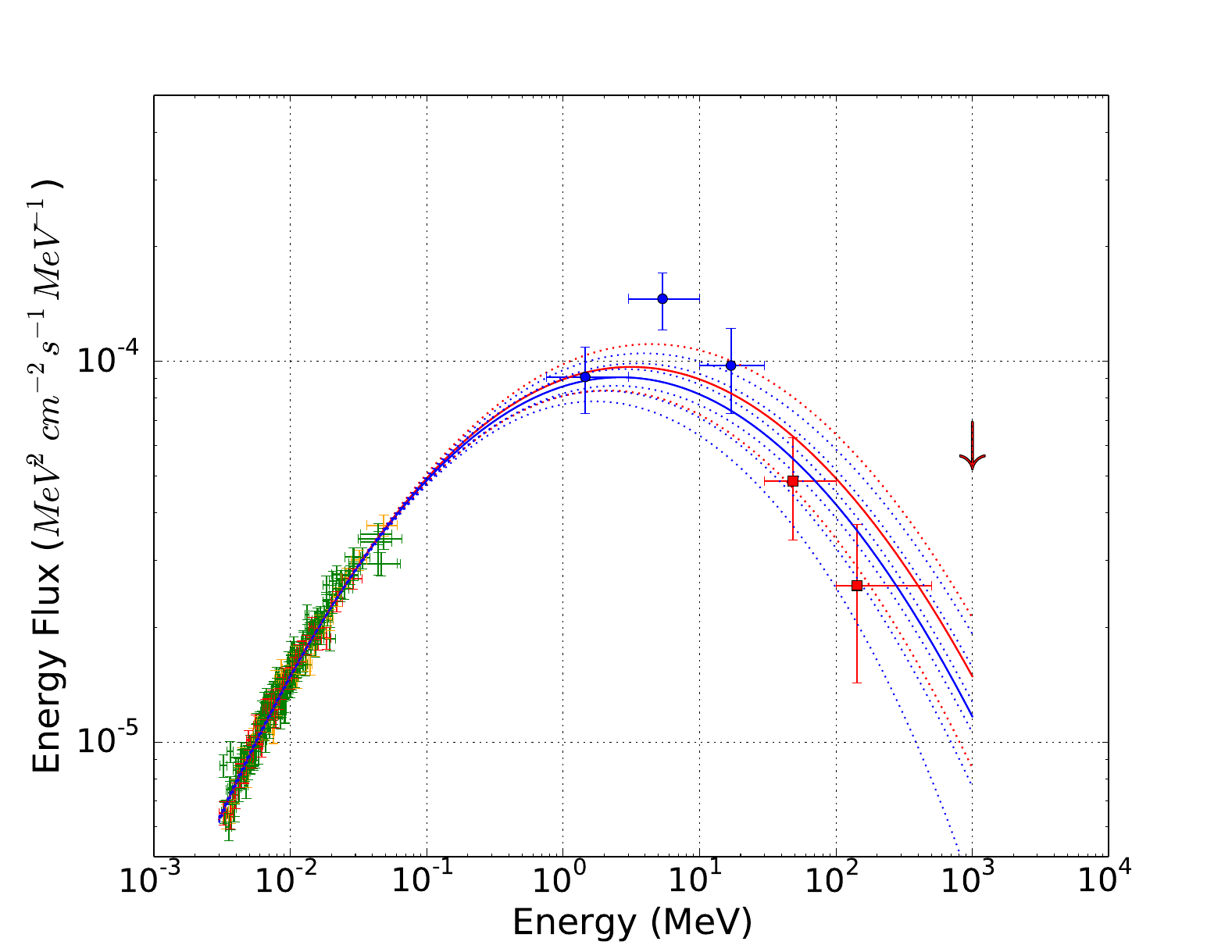}
\caption{The pulsed broadband energy spectrum of PSR B1509$-$58 from soft X-ray to $\gamma$-ray. The red squares are the {\em AGILE} observations in the 30--100 MeV and 100--500 MeV bands (\citealt{2010ApJ...723..707P}). The arrow is the {\em AGILE} upper limit above 500 MeV band (\citealt{2010ApJ...723..707P}). The blue circles are the {\textsc COMPTEL} observations in the 0.75--3 MeV, 3--10 MeV and 10--30 MeV bands (\citealt{1999A&A...351..119K}). Data from 3-79 keV are from our {\em NuSTAR} observations. The solid red line is the logarithmic parabolic model obtained from {\em NuSTAR} data, and the dashed red lines show its $1\sigma$ range. The solid blue line is the logpar model obtained from all data from 3 keV to 500 MeV, and the dashed blue lines are its 1, 2 and 3$\sigma$ confidence regions.
\label{fig:broad}
}
\end{figure}

\subsection{Broadband Spectrum}
\label{sec:broad}
We also produced a 3 keV-- 500 MeV broadband spectrum for PSR B1509$-$58 and show it in Figure \ref{fig:broad}. The X-ray data are from {\em NuSTAR} and the $\gamma$-ray data are from observations made with COMPTEL (\citealt{1999A&A...351..119K}) and {\em AGILE} (\citealt{2010ApJ...723..707P}). Though significant pulsations were also detected by \textit{Fermi} LAT (30-- 1000 MeV) (den Hartog et al. in preparation), we did not use the results for our spectral analysis, because the \textit{Fermi} SED of PSR B1509$-58$ was rather uncertain. Table \ref{ta:broadband} shows results obtained in 3--79 keV, 3 keV--30 MeV and 3 keV--500 MeV by fitting the logpar model (Equation \ref{eq:logpar}) with data from {\em NuSTAR}, {\em NuSTAR} \& {\textsc COMPTEL}, and {\em NuSTAR} \& {\textsc COMPTEL} \& {\em AGILE}, respectively. All fittings were acceptable. Note that we have unfolded the spectra in two slightly different methods and show the corresponding results in Tables \ref{ta:pulsed} and \ref{ta:broadband}, respectively. These two methods are compatible, since in the 3-79 keV band they produce results that are consistent within 2$\sigma$  (the second line in Tables \ref{ta:pulsed} and the first line in Table \ref{ta:broadband}). The first method removes well the effects from the observations' Ancillary Response Files (ARFs) and Response Matrix Files (RMFs), but required data with response files. In the broadband spectral analysis, since the response files of the $\gamma$-ray data were unavailable to us, we unfolded spectra with a different method assuming that the RMFs were perfectly diagonal, which is generally a fair approximation.
As shown in Table \ref{ta:broadband}, spectral parameters obtained in these three energy intervals were consistent with each other within 1$\sigma$, hence the energy spectrum from 3 keV to 500 MeV can be represented with one smooth logarithmic parabolic model. Figure \ref{fig:broad} extrapolates the model obtained from {\em NuSTAR} alone (the red solid line) and its 1$\sigma$ ranges (the dashed red lines) to the $\gamma$-ray band.  The solid blue line is the model obtained from a combined fit of all data from 3 keV to 500 MeV (see Table 4) and the dashed blue lines are its 1, 2 and 3$\sigma$ ranges. 

Table \ref{ta:broadband} also shows results from \citet{2001A&A...375..397C} in two energy bands. Results in 2-300 keV were obtained from {\em BeppoSAX} and broadband results in 2 keV to 30 MeV were from a combined fit of {\em BeppoSAX} and {\textsc COMPTEL}. Their broadband spectral parameters $\alpha$ and $\beta$ are consistent within 1.5$\sigma$ with ours (3 keV to 500 MeV), but their normalization parameter $K$ is $\sim$4.7 $\sigma$ smaller than ours. In this paper, when comparing two results, $\sigma$ refers to the root-mean-square of uncertainties on the two values unless stated otherwise.


\subsection{Phase-resolved and Pulsed Spectral Analysis}
\label{resolvedspecana}

\citet{2012ApJS..199...32G} obtained phase-resolved X-ray spectra of PSR B1509$-$58 with the {\em Rossi X-Ray Timing Explorer (RXTE)} from 3--30 keV and reported variations of the phase-resolved photon indices. We used the {\em NuSTAR} observation to extend this analysis up to 79 keV and compare to their results. Moreover, we fitted the spectra with both the simple power-law and the logpar models, and compared the fits in each phase interval.

\begin{figure}[]
\includegraphics[width=0.5\textwidth]{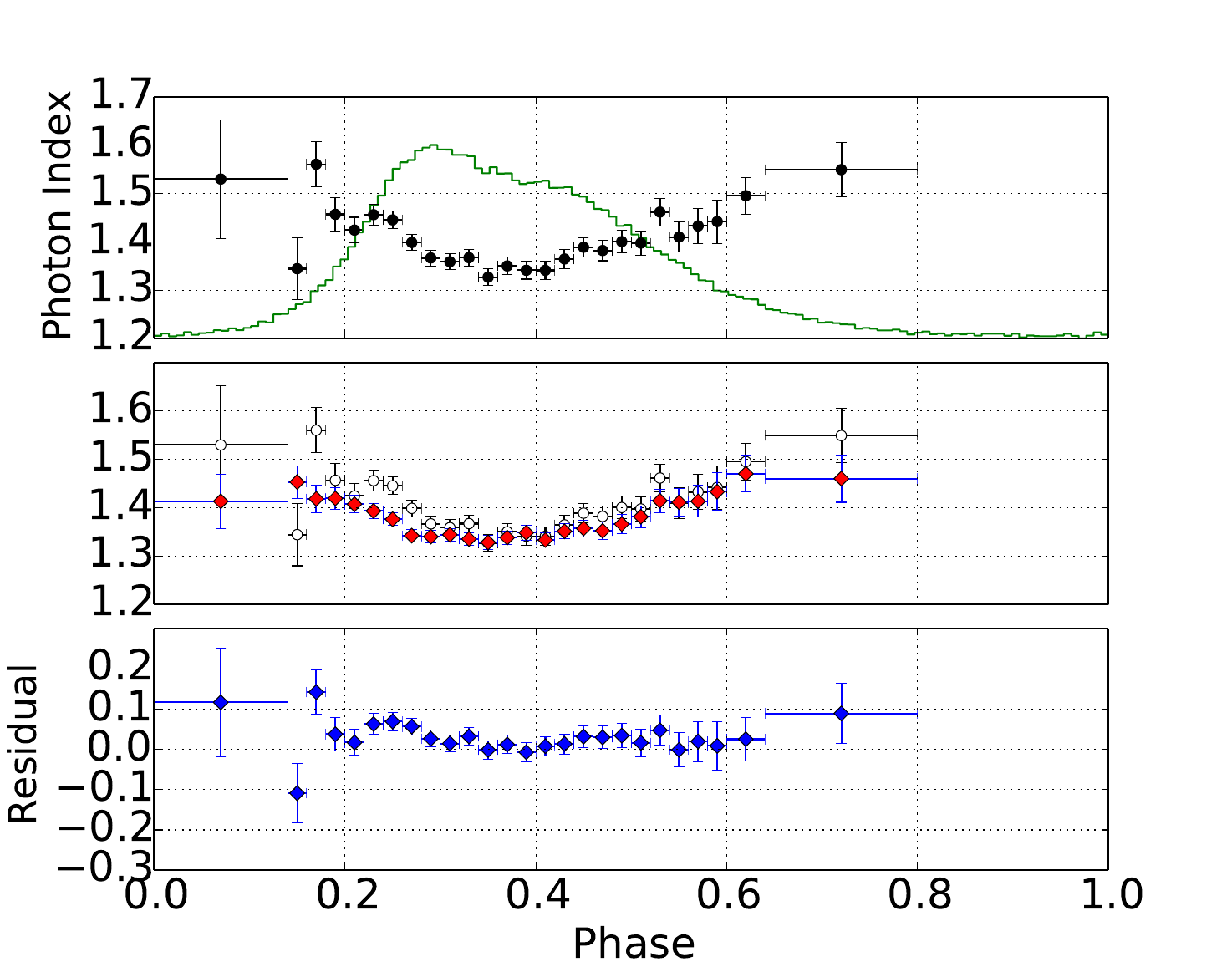}
\caption{Phase-resolved and pulsed spectral fits with the power law model. Top panel: photon indices of PSR B1509$-$58 at different phase ranges (black filled circles). The green line illustrates the pulse profile. Middle: comparison of the photon indices obtained in this paper (open black circles) with those of Ge et al (red filled diamonds). Bottom: difference between the photon indices obtained in this paper and in \citet{2012ApJS..199...32G}.
\label{fig:resolved}}
\end{figure}

We performed a phase-resolved and pulsed spectral analysis for 26 phase bins in the energy range of 3--79 keV. For this analysis, the pulsar events were extracted with a circular region of aperture radius 60$''$ centered on the pulsar position (Section \ref{sec:timingana}). 
To compare with the results of \citet{2012ApJS..199...32G}, we used the same off-pulse interval as they did. The pulse profile in the 3--79 keV band is illustrated in the top panel of Figure \ref{fig:resolved}. We defined phase interval 0 to 0.8 as the pulsed portion and the mean count rate between phases 0.8 to 1.0 as the off-pulse level. The off-pulse level was then subtracted from the pulsed portion, yielding the pulsed spectrum. Then each pulsed spectrum was divided into 26 phase bins to produce phase-resolved and pulsed spectra. Each phase interval had at least 5700 counts in 3-79 keV. To compare results with Ge et al. (2012; see  Fig. \ref{fig:resolved} middle panel), we chose the same phase zero as they did, and divided phase bins using the same phase values. 
Each spectrum was grouped into at least 20 counts per bin and then fitted jointly to the power-law model using {\em XSPEC}. We checked the results with both $\chi^2$ minimization and the {\ttfamily cstat} statistic. The results did not differ significantly, so we only report the results from $\chi^2$ statistics. For reasons stated in Section \ref{sec:spectrumana}, we fixed $N_{\rm H}$ at $0.95\times 10^{22}\rm \ cm^{-2}$ and tied all the fitting parameters except the cross normalization factors of the eight spectra in each phase bin.

The power-law model is acceptable in all phase intervals. Table \ref{ta:resolved} lists the results and Figure \ref{fig:resolved} shows the photon indices as a function of phase. The photon indices decrease from about 1.53 to 1.34 within the rising phase interval, keep stable around 1.34 between phases 0.28 to 0.44, and start to increase again around phase 0.54. As shown in the middle and bottom panels of Figure \ref{fig:resolved}, all results are consistent within 3$\sigma$ with those of \citet{2012ApJS..199...32G} obtained by {\em RXTE} in the 3--30 keV band. 

\begin{figure}[]
\includegraphics[width=0.5\textwidth]{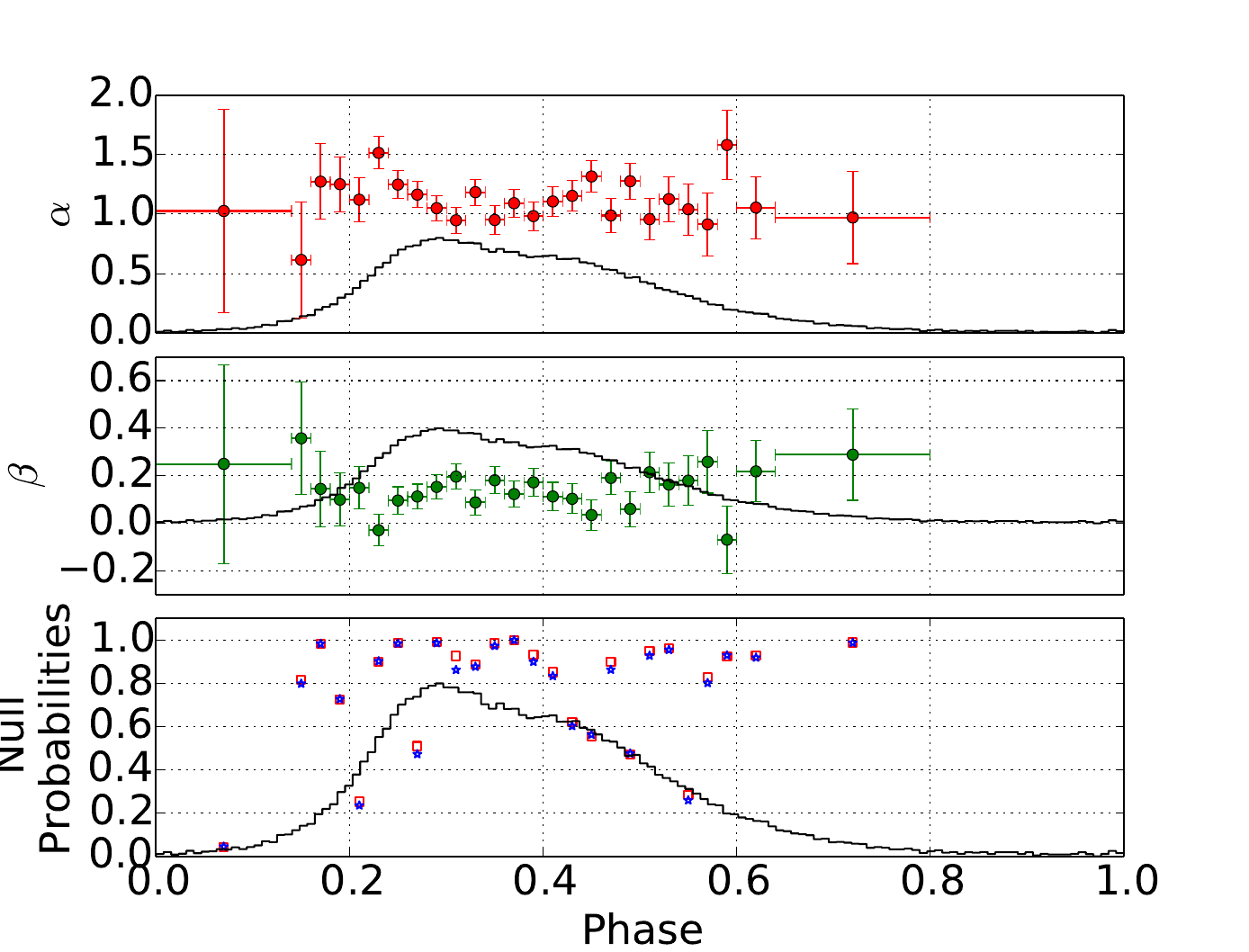}
\caption{Phase-resolved and pulsed spectral fits with the logpar model (Equation 1). The black line in each panels shows the pulse profile. Top panel: The parameter $\alpha$ as a function of phase. Middle: The parameter $\beta$ as a function of phase. Bottom: Null hypothesis probabilities obtained from $\chi^2$ tests for the power-law model (red open squares) and the logpar model (blue open stars). Many of them are very close to each other.
\label{fig:resolved_logpar}}
\end{figure}

The logpar model is also acceptable in all phase intervals. Table \ref{ta:resolved} lists the results and Figure \ref{fig:resolved_logpar} shows the parameters as a function of phase. Unlike the photon index of the power-law model, the logpar model parameter $\alpha$ does not have an obvious evolving trend within the phase interval $\sim$0--0.22, but decreases in the phase interval 0.22--0.3, keeps stable after the phase peak within the phase interval of $\sim$0.3--0.44, and shows no obvious trend after phase 0.44 (Fig. \ref{fig:resolved_logpar} top panel). The middle panel of Figure \ref{fig:resolved_logpar} shows the parameter $\beta$ as a function of phase. Note that the $\beta$ values are negative in phase intervals 0.22--0.24 and 0.58--0.6, which means that the spectra become harder in those two phase intervals. In the other 24 phase intervals, $\beta$ values are positive, which implies that the spectra become softer. A larger positive $\beta$ value indicates that the spectra become soft faster in that phase interval, but our parameter $\beta$ does not show an obvious evolving trend with phase.

To compare the power-law model to the logpar model at different phase intervals, we listed their null hypothesis probabilities obtained from the $\chi^2$ minimization tests in Table \ref{ta:resolved}, and plotted them in the bottom panel of Figure \ref{fig:resolved_logpar}. Both models were acceptable in all phase intervals, and in 20 out of the 26 phase intervals the null hypothesis probabilities were above 68\%. The logpar model did not make a significant improvement compared to the power law model in any of the 26 phase intervals. The differences between the null hypothesis probabilities of the two models were less than 4\% in 25 out of the 26 phase intervals, and the largest improvement in null hypothesis probability was 6.4\%, which is insignificant. In addition, the bottom panel of Figure \ref{fig:resolved_logpar} illustrates that the null hypothesis probabilities for neither model has any obvious correlation with the phase values. 

\begin{table*}
\begin{center}
\caption{The Phase-resolved and Pulsed Spectral Analysis Results}
\label{ta:resolved}
\begin{tabular}{c|cccc|ccccc}
\hline\hline
 & \multicolumn{4}{c}{Power Law} & \multicolumn{5}{|c}{Logpar}\\
 \cline{2-10}
Phase &  Photon Index  &  $\chi^2$ & dof & Null Hypothesis&  $\alpha$  & $\beta$ & $\chi^2$ & dof & Null Hypothesis \\
& & & & Probability & &&&& Probability \\
\hline
0.00-0.14 & 1.530$\pm$0.123 & 1207.66 &1126  &0.045& 1.026$\pm0.852$ & $0.250\pm0.418$ &1207.35 & 1125  & 0.044  \\
0.14-0.16 & 1.344$\pm$0.064 & 264.861 & 285  &0.798&  $0.615\pm0.488$ & $0.357\pm0.235$ & 262.49 &284  & 0.815 \\
0.16-0.18 & 1.560$\pm$0.046& 297.938 &352    &0.983& $1.273\pm$0.318 & $0.145\pm0.158$ &297.09 &351    & 0.983 \\
0.18-0.20 & 1.456$\pm$0.034 & 417.727 & 436  &0.727& $1.251\pm0.232$ & $0.100\pm0.112$ & 416.925 & 435  & 0.725 \\
0.20-0.22 & 1.424$\pm$0.026 & 573.546 & 550  &0.236& $1.121\pm0.182$ & $0.149\pm0.089$ & 570.705 & 549  & 0.252 \\
0.22-0.24 & 1.456$\pm$0.022 & 634.603 & 682  &0.902& $1.515\pm0.136$ & $-0.029\pm0.066$ &634.411 &681   & 0.899 \\
0.24-0.26 & 1.445$\pm$0.018 & 720.945 & 805  &0.984& $1.248\pm0.119$ & $0.096\pm0.058$ & 718.06 &804   & 0.986  \\
0.26-0.28 & 1.398$\pm$0.017 & 874.19 & 872   &0.473& $1.164\pm 0.109$ & $0.113\pm0.052$ & 869.266 & 871 & 0.510 \\
0.28-0.30 & 1.366$\pm$0.016 & 813.851 & 905  &0.986& $1.050\pm0.108$ & $0.153\pm0.052$ & 804.684 & 904  & 0.992 \\
0.30-0.32 & 1.358$\pm$0.016 & 843.144 & 889  &0.862& $0.948\pm0.109$ & $0.196\pm$0.052 & 827.622 & 888  & 0.926 \\
0.32-0.34 & 1.367$\pm$0.017 & 825.724 & 874  &0.877& $1.184\pm$0.110 & $0.088 \pm0.053$ & 822.84 & 873  & 0.886   \\
0.34-0.36 & 1.327$\pm$0.018 & 741.81 & 818   &0.973& $0.951\pm0.119$ & $0.181\pm$0.057 &730.939 & 817  & 0.986   \\
0.36-0.38 & 1.350$\pm$0.018 & 669.523 & 812  &1.000& $1.091\pm0.117$ & $0.123\pm0.055$ &664.362 & 811   & 1.000 \\
0.38-0.40 & 1.341$\pm$0.018 & 728.888 & 779  &0.900& $0.983\pm0.121$ & $0.172\pm0.058$ & 719.396 & 778  & 0.934\\
0.40-0.42 & 1.341$\pm$0.019 & 735.784 & 774  &0.834& $1.106\pm0.125$ & $0.113\pm0.060$ &732.077 & 773  & 0.851 \\
0.42-0.44 & 1.364$\pm$0.019 & 747.241 & 758  &0.603& $1.153\pm0.130 $ & $0.103\pm0.062$ & 744.481 & 757 & 0.620\\
0.44-0.46 & 1.388$\pm$0.020 & 718.338 & 725  &0.563& $1.316\pm0.132$ & $0.035\pm0.063$ & 718.03 & 724  & 0.556  \\   
0.46-0.48 & 1.382$\pm$0.021 & 635.89 & 676   &0.863& $0.988\pm0.145$ & $0.191\pm0.070$ & 628.223 & 675  & 0.901 \\
0.48-0.50 & 1.400$\pm$0.023 & 618.462 & 617  &0.476& $1.276 \pm0.154$ & $0.060\pm0.074$ & 617.793 & 616 & 0.472\\ 
0.50-0.52 & 1.397$\pm$0.025 & 516.611 & 565  &0.928& $0.957\pm0.176$ & $0.215\pm0.086 $ & 510.051 &564  & 0.949\\   
0.52-0.54 & 1.461$\pm$0.028 & 459.932 & 513  &0.955& $1.126\pm$0.189 & $0.163\pm0.092$ & 456.378 &512  & 0.963 \\   
0.54-0.56 & 1.410$\pm$0.032 & 483.215 & 464  &0.260& $1.040\pm0.216$ & $0.180\pm0.104 $ & 480.017 &463  & 0.283\\
0.56-0.58 & 1.432$\pm$0.037 & 385.549 & 410  &0.802& $0.913\pm0.267$ & $ 0.258\pm0.131$ & 381.653 &409  & 0.830\\   
0.58-0.60 & 1.442$\pm$0.045 & 322.192 & 361  &0.930& $1.581\pm 0.294$ & $-0.069\pm0.142$ & 321.971 &360 & 0.926\\   
0.60-0.64 & 1.495$\pm$0.038 & 565.61 & 614   &0.919& $1.055\pm0.262$ & $0.218\pm0.128$ &562.616 &613   & 0.928 \\
0.64-0.80 & 1.549$\pm$0.056 & 1306.21 & 1424 &0.988& $0.972\pm0.388$ & $0.288\pm 0.192$ & 1303.88 &1423 & 0.989\\
\hline\hline
\end{tabular}
\end{center}
\end{table*}

\section{Discussion}
\label{sec:disc}
We have reported on a pulsed spectral analysis and timing analysis of PSR B1509$-$58 with observations made by {\em NuSTAR} and {\em Chandra}, as well as previous results measured by COMPTEL and {\em AGILE}. The phase-averaged and pulsed spectral analysis shows that a smooth logarithmic parabolic model represents our 3-79 keV {\em NuSTAR} data better than 
a simple power-law model (Fig. \ref{fig:models} and Table \ref{ta:pulsed}). In addition, the logpar model describes our broadband spectrum well from 3 keV up to 500 MeV (Fig. \ref{fig:broad} and Table \ref{ta:broadband}). The phase-resolved and pulsed spectral analysis shows that the photon indices in 26 phase intervals are consistent with the results of \cite{2012ApJS..199...32G} (Fig. \ref{fig:resolved} and Table \ref{ta:resolved}). However, the logpar model does not describe our spectra significantly better than the power-law model in any of these phase intervals (Fig. \ref{fig:resolved_logpar}). In the timing analysis, we found that the pulsed profiles in 14 energy bands between 3-79 keV show no significant differences (Fig. \ref{fig:pp}). We measured the pulsed fraction of PSR B1509$-$58 in the hard X-ray band for the first time (Fig. \ref{fig:PF}). The measured pulsed fraction increased from $\sim$0.68 to $\sim$0.88 from 3 to 19 keV because of the PWN contamination which is significant due to the broad {\em NuSTAR} PSF, and then reached a plateau near the latter value within 19--79 keV. 
\subsection{Pulsed and Phase-Averaged Spectrum}
\label{disc_spec}



\begin{figure}
\centering
\includegraphics[width=0.5\textwidth]{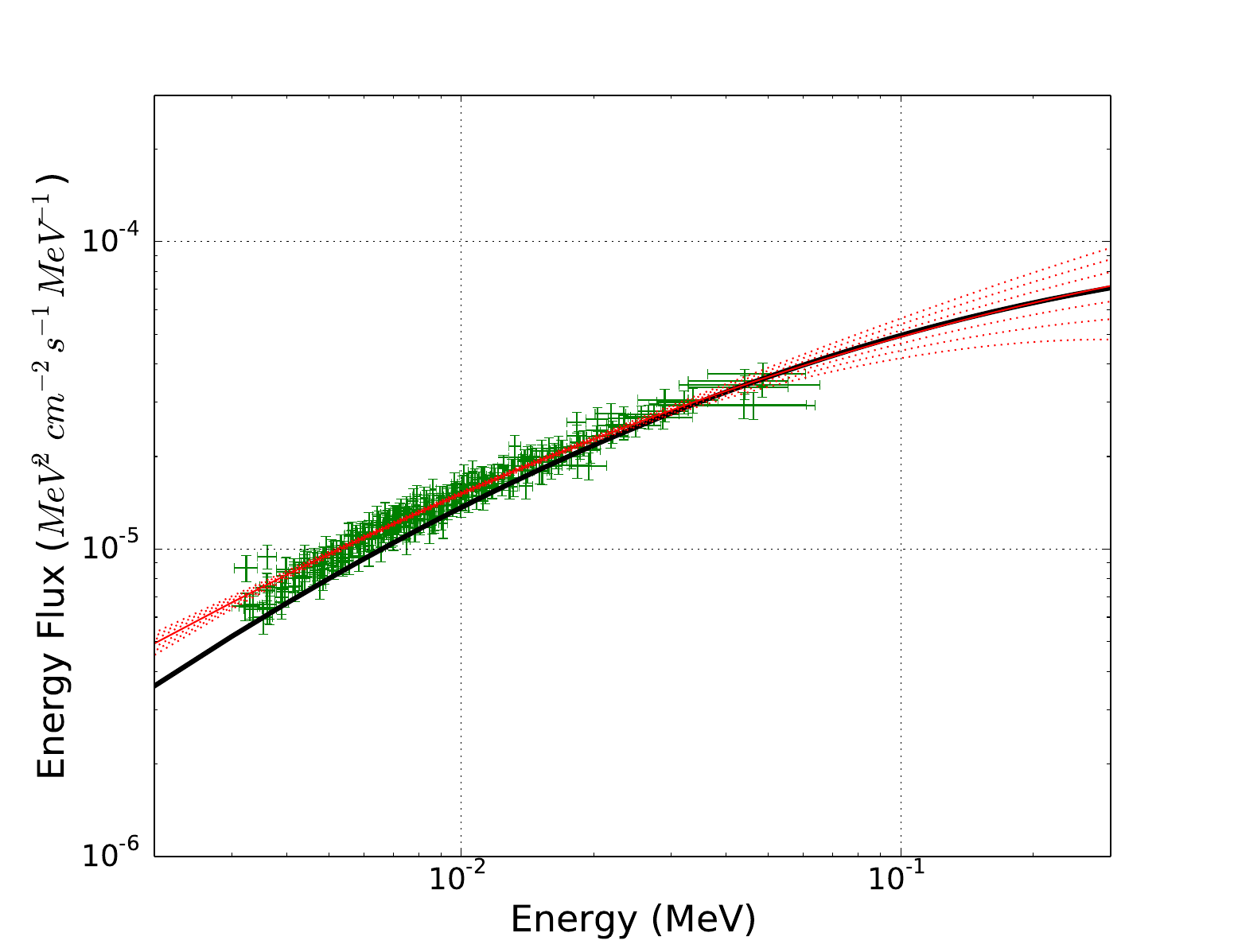}
\caption{Energy spectrum of our {\em NuSTAR} observation and that from \citet{2001A&A...375..397C}. The green points are {\em NuSTAR} data. The solid black curve is the result of \citet{2001A&A...375..397C}. The solid red curve is the {\em NuSTAR} logpar spectral model, and the dotted curves are its 1, 2 and 3$\sigma$  ranges. 
\label{fig:x-ray_cusunamo}
}
\end{figure}

In comparison to the results of \citet{2001A&A...375..397C} (Table \ref{ta:pulsed}), uncertainties in our parameters are 50\%-75\% smaller. To compare the shape of our logpar model to Cusumano's, we fixed all parameters at Cusumano's values but let the normalization factors vary, and fitted our {\em NuSTAR} data with this model in {\em XSPEC} with $\chi^2$ minimization. The results were not acceptable, with a probability smaller than 10$^{-8}$. Figure \ref{fig:x-ray_cusunamo} compares Cusumano's model to our logpar model, and to the {\em NuSTAR} data. The difference in shape is visible. Moreover, our pulsed flux at 2--10 keV is ($2.39\pm0.04)\times10^{-11}$ erg cm$^{-2}$ s$^{-1}$, significantly larger than the $2.0\times10^{-11}$ erg cm$^{-2}$ s$^{-1}$ reported by \citet{2001A&A...375..397C}, but consistent with the $(2.30\pm0.02)\times10^{-11}$erg cm$^{-2}$ s$^{-1}$ reported by \citet{2012ApJS..199...32G}, and the $(2.52\pm0.05) \times10^{-11} \rm erg\rm~cm^{-2}\rm~s^{-1}$ reported by \cite{2015MNRAS.449.3827K}. However, Cusumano's results were determined from {\em BeppoSAX} observations. As argued in Section \ref{sec:intro}, we consider our {\em NuSTAR} results more reliable than the {\em BeppoSAX} results, as {\em BeppoSAX} had systematic cross-calibration discrepancies among its three instruments among different energy bands.

We also compared the results of our observations with previous X-ray observations and summarize the comparison in Table \ref{ta:x-ray}. A simple extrapolation of the observation results of {\em OSSE} (50 keV--5 MeV), {\em Ginga} (3--60 keV), {\em RXTE} (2--250 keV) and {\em WELCOME} (94--240 keV) into our energy range predicts a pulsed flux of $\sim$1.26$\times10^{-10}$, $\sim$1.50$\times10^{-10}$, $\sim$2.20$\times10^{-10}$ and $\sim$3.14$\times10^{-10}$ erg cm$^{-2}$ s$^{-1}$, respectively. The  pulsed fluxes predicted by {\em OSSE} and {\em Ginga} are consistent with our {\em NuSTAR} flux in 3--79 keV, while {\em RXTE} and {\em WELCOME}  each predict a pulsed flux of $\sim$2 and $\sim$3 times larger than our {\em NuSTAR} flux. Note that cross-calibration discrepancies among {\em NuSTAR}, {\em Ginga} and {\em RXTE} against the Crab nebula have been found. {\em NuSTAR} was reported to measure the Crab spectrum $\sim$15\% lower than {\em Ginga} and {\rm RXTE} (\citealt{madsen2015calibration}; \citealt{2005SPIE.5898...22K}). Also {\em NuSTAR} has an uncertainty of $\sim$5\% on its absolute flux measurement, and the Crab flux could change by a few percent per year (\citealt{2041-8205-727-2-L40}). 
Furthermore, the {\em Ginga} (3--60 keV), {\em RXTE} (2--250 keV), {\em WELCOME} (94--240 keV) and {\em OSSE} (50 keV--5 MeV) observations were reported to fit with simple power-law models. Their photon indices were $1.30\pm0.05$ ({\em Ginga}; \citealt{1993AIPC..280..213K}), $1.358\pm0.014$ ({\em RXTE}; \citealt{marsden1997x}), $1.64^{+0.43}_{-0.42}$ ({\em WELCOME};  \citealt{1994ApJ...428..284G}) and $1.68\pm0.09$ ({\em OSSE};  \citealt{1994ApJ...434..288M}) (Table \ref{ta:x-ray}). This indicates that the pulsed spectrum becomes softer as energy increases, consistent with the behavior of our logpar model. 

\begin{table}
\begin{center}
\caption{Predicted pulsed flux}
\label{ta:x-ray}
\begin{tabular}{cccc}
\hline\hline
 Observation & Energy band & Photon Index & Flux in 3--79 keV \\
 & [keV] & & [10$^{-10}$erg cm$^{-2}$ s$^{-1}$] \\
\hline
{\em NuSTAR} & 3--79 & $1.386\pm0.007$ & $1.136\pm0.018$ \\
\hline
{\em Ginga} & 3--60 & $1.30\pm0.05$ & $\sim$1.50 \\
\hline
{\em RXTE} & 2--250 & $1.358\pm0.014$ & $\sim$2.20 \\
\hline
{\em WELCOME} & 94--240 & $1.64^{+0.43}_{-0.42}$ & $\sim$3.14 \\
\hline
{\em OSSE} & 50--5000 & $1.68\pm0.09$ & $\sim$1.26 \\
\hline\hline
\end{tabular}
\end{center}
\end{table}

\begin{figure}
\centering
\includegraphics[width=0.5\textwidth]{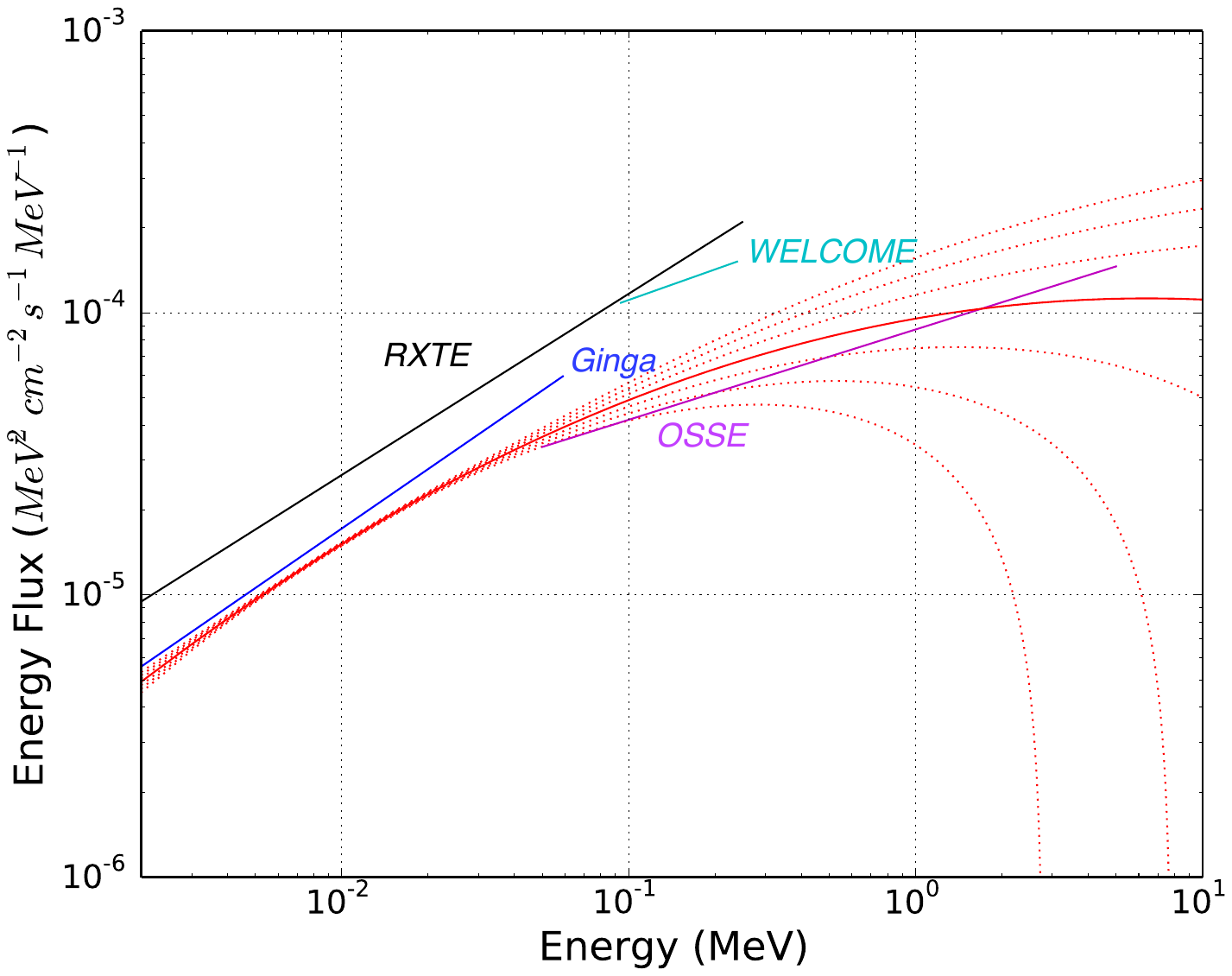}
\caption{Energy spectrum of {\em NuSTAR} data and previous X-ray observations. The solid red line is the {\em NuSTAR} logpar spectral model, and the dotted red lines are its 1, 2 and 3$\sigma$ ranges. The solid black line, solid cyan line, solid blue line, and solid magenta line, from top to bottom, are X-ray spectra observed by {\em RXTE}, {\em WELCOME}, {\em Ginga} and {\em OSSE}, respectively.
\label{fig:x-ray}
}
\end{figure}

Figure \ref{fig:x-ray} shows a comparison of the best-fit energy spectrum from our {\em NuSTAR} observations with those of the four previous X-ray observations. {\em OSSE} is within 3$\sigma$ from our {\em NuSTAR} spectrum, while {\em RXTE}, {\em WELCOME} and {\em Ginga} do not agree well with our new {\it NuSTAR} measurements. This perhaps is due to the cross-calibration issues between instruments described earlier in this section, since the source flux has been very stable over a long period (\citealt{2011ApJ...742...31L}). Note that the uncertainty of the model increases dramatically with energy.

In their broadband spectral analysis, \citet{2001A&A...375..397C} claimed that a smooth logarithmic parabolic model could represent the spectrum well up to 30 MeV, but must break more rapidly afterwards. However, our result shows that a logarithmic parabolic function can represent the spectrum well up to 500 MeV. 
Moreover, one interesting feature of the broadband energy spectrum is its peak energy. According to Equation \ref{eq:logpar}, the energy flux should have a maximum at the following energy: 
\begin{equation} \label{eq:max}
E_m = E_0 10^{[(2 - \alpha)/(2 \beta)]},
\end{equation}
which corresponds to 
3 MeV and a 90\% upper limit of 8 MeV, using the best-fit parameters obtained from {\em NuSTAR} alone, $4\pm2$ MeV using the combined fit of {\em NuSTAR} and COMPTEL from 3 keV to 30 MeV, and $2.6\pm0.8$ MeV using results obtained from all data up to 500 MeV. This value is consistent with the recent results of \cite{2015MNRAS.449.3827K}. Using multi-year data across $\sim$2 keV--1 GeV from \textit{RXTE PCA/HEXTE}, \textit{Fermi LAT}, \textit{COMPTEL} and \textit{INTEGRAL ISGRI}, they performed a broad-band spectral fit of the pulsed emission of PSR B1509$-$58 to a power-law model with a modified exponential cutoff, and derived a maximum luminosity per energy decade at $\sim$2.5 MeV.
The result of \cite{2001A&A...375..397C} was obtained from a combined fit of their {\em BeppoSAX} X-ray observation and COMPTEL, from 1 keV to 30 MeV. Their peak energy was
5 MeV, with a 90\% upper limit of 14 MeV. Their value agrees with all of ours, though their uncertainty is $\sim$9 times larger than what we have obtained from the combined fit. 
Note that each of the $\gamma$-ray data points plays an important role in our combined fits, and in the determination of the peak value.
 
\citet{2001A&A...375..397C} also discussed the relation of the peak values between PSR B1509$-$58 and the Crab pulsar. For the outer gap model, the synchrotron spectrum emitted from the secondary electron-positron pairs created by inward flowing curvature photons of Crab-like pulsars is predicted to obey the following scaling relation (\citealt{2001A&A...375..397C}):
\begin{equation}\label{eq:scaling}
(E_m/E_{mC}) \sim (P/P_C)^{7/4}(\dot{P}/\dot{P_C})^{1/4}.
\end{equation}
Here, $P_C$ and $\dot{P_C}$ are period and spin-down rate of the Crab, $E_m$, $E_{mC}$ are energies corresponding to the spectral peaks of the Crab-like pulsar and the Crab. It is suggested that a Crab-like pulsar spectrum that follows this relation is likely to be synchrotron radiation from the secondary pairs (\citealt{2001A&A...375..397C}). For PSR B1509$-$58, the RHS of Equation \ref{eq:scaling} is $\sim$20. For the Crab, the SED peak energy ($E_{mC}$) is $\sim$14 keV for the main pulse, and $\sim$200 keV for the interpulse component.
\citet{2001A&A...375..397C} concluded that their observation agreed with this ratio when the spectrum of the Crab's interpulse region was considered, but it was about one order of magnitude larger than the result when the spectrum of the Crab's main pulse was considered. We can improve on this result thanks to our more precisely measured peak energy. From our analysis, 
the peak value obtained from our combined fit (3 keV--500 MeV) gives a ratio of $\sim$$5\pm2$ when considering the Crab's interpulse region, and $\sim$$80\pm20$ for the Crab's main pulse. Neither agrees well with the ratio of $\sim$20 predicted by Equation \ref{eq:scaling}. Note that the exact uncertainties of $E_{mC}$ or the RHS of Equation \ref{eq:scaling} are unavailable to us, so the conclusion is based on estimations.


\subsection{Predicted Spectra from Theoretical Models}
PSR B1509$-$58 has an atypical spectrum in the X-ray and $\gamma$-ray bands. $\gamma$-ray pulsars typically have spectral peaks in the GeV energy band (\citealt{2004ASSL..304..149T}), while PSR B1509$-$58 was observed to emit very little $\gamma$-ray emission, and its energy spectrum peaks in the hard X-ray band at $2.6\pm0.8$ MeV (Section \ref{sec:broad}). \citet{0004-637X-764-1-51} proposed a model in the context of outer gap acceleration to explain the weak $\gamma$-ray emission from this pulsar and calculated the X-ray spectrum. They proposed that the outgoing GeV--band $\gamma$-ray radiated via curvature radiation from the electron-positron pairs created in the null charge surface are missed because of the Earth viewing angle, while the inward propagating $\gamma$-rays passing near the pulsar surface may be absorbed by the magnetic field and converted into electron-positron pairs. The secondary pairs that travel with sufficiently large pitch angles with respect to the magnetic field lines emit synchrotron radiation which peaks at $\sim$10 MeV, and covers a wider sky area compared with curvature radiation of the primary particles. They concluded that only synchrotron radiation could be observed due to our viewing angles. However, the model may under-predict emissions in the X-ray band. For example, \citet{0004-637X-764-1-51} compared the energy spectrum predicted by their model with previous X-ray observations by {\em Ginga}, {\em OSSE}, {\em WELCOME} and {\em RXTE}, and found that the model predicts lower energy flux than is seen in all these observations except that of {\em OSSE}. The predicted  energy flux is $\sim$half as much as observed by {\em RXTE} and {\em WELCOME}, and more than 5$\sigma$ lower than the {\em RXTE} observation.

We quantitatively compared our results with the theoretical model prediction that was shown in Figure 6 of \citet{0004-637X-764-1-51}. By inspection, their theoretical prediction is roughly consistent with our {\em NuSTAR} 3$\sigma$ ranges (Fig. \ref{fig:x-ray}). Moreover, the theoretical model predicts a peak at $\sim$3 MeV (\citealt{0004-637X-764-1-51}), which agrees with our results.

\subsection{Pulsed and Phase-resolved Spectrum}

We also studied the pulsed, phase-resolved spectrum of PSR B1509$-$58.  We found that the photon index is $1.34 \pm 0.02$ in the middle of the pulse, becoming larger at the two sides, for the power-law model. Our results are largely consistent with those of \citet{2012ApJS..199...32G}. Note that in 24 phases our photon indices are slightly larger than those of \citet{2012ApJS..199...32G} (see Fig. \ref{fig:resolved}, middle and bottom panels). This may be consistent with the fact that the photon index increases with energy in the X-ray band, since our results were obtained from {\em NuSTAR} between 3--79 keV, higher than their 3--30 keV energy band. 

\citet{2012ApJS..199...32G} performed pulsed and phase-resolved spectral analyses for three young and bright X-ray pulsars: the Crab pulsar, PSR B1509$-$58 and PSR B0540$-$69. Their results show that the photon indices of PSRs B1509$-$58 and B0540$-$69 are harder at the centres of the pulse profiles and softer at the wings, while those of the Crab pulsar are softer at its two peaks. However, their timing analysis shows that the pulse profiles of PSRs B1509$-$58 and B0540$-$69 could each be further decomposed into two narrower Gaussian components. Taking this into account, they concluded that overall the three pulsars' photon indices had a similar evolving trend with their pulsed flux. Our {\em NuSTAR} observations support the same conclusion. 

 
\subsection{Pulsed Fraction}
\label{disc_PF}


\cite{0004-637X-764-1-51}'s model clearly predicted X-ray off-pulse emission from the pulsar, and those authors suggested that it was probably produced by electrons accelerated transverse to the field line. We have measured the pulsed fraction of PSR B1509$-$58 with {\em NuSTAR} in the 3--79 keV band and with {\em Chandra} HRC in the 0.5--10 keV band. The observed off-pulse emission originated from both the pulsar and the PWN. With our pulsed fraction measurements, we can determine the contributions from the PWN and the PWN-free pulsar off-pulse components to the total emission, and estimate the photon index of the pulsar's off-pulse component for the first time.

Based on the results shown in Figure \ref{fig:PF}, 
we had that at 3-4 keV, the PWN-free  pulsed fraction measured with {\em Chandra} is:
$$PF_{\rm PWN-free}=\frac{pulsed}{pulsed+offpulse_{\rm PWN-free}} = 0.88\pm0.06,$$
while the PWN-included result from {\em NuSTAR} is:
\begin{equation}
\begin{split}
\notag PF_{\rm withPWN} & = \frac{pulsed}{pulsed+{offpulse}_{\rm PWN-free}+PWN} \\&= 0.68\pm0.04.
\end{split}
\end{equation}
From these two results, we derived that the PWN contrioff-pulsebutes $0.23\pm0.06$ of the total number of {\it NuSTAR} photons, and that the pulsar's PWN-free off-pulse component contributes $0.09\pm0.07$ of the total number of photons. The latter result is roughly consistent with what is shown in Figure \ref{fig:PF}, where the pulsed fraction approaches 90\%. The number of photons detected from the PWN is $3\pm2$ times of those from the pulsar's PWN-free off-pulse component at $\sim$3--4 keV. Consequently, the pulse-off components at soft X-ray bands are likely to be PWN-dominated, and as energy increases the PWN contribution decreases until $\sim$19 keV, where it no longer matters. The off-pulse components of 19--79 keV are composed of the pulsar's PWN-free off-pulse component alone.

Additionally, we can estimate the photon index of the pulsar's PWN-free off-pulse component. The PWN spectrum is much softer than the pulsar's: the photon indices of the PWN and the pulsar are 1.6--1.8 (\citealt{an2014high}) and $1.386\pm0.007$, respectively. The PWN is also much fainter than the pulsar: at 19 keV, its flux is less than 1\% of the pulsar's.
Hence we can safely neglect the PWN contribution above 19 keV and regard the {\em NuSTAR} results between 19--25 keV and 25--79 keV as PWN-free pulsed fractions of the pulsar. Meanwhile, the {\em Chandra} HRC results also give the PWN-free pulsed fraction. Figure \ref{fig:PF} shows that the four PWN-free pulsed fractions, namely the two {\em NuSTAR} results in 19--25 keV and 25--79 keV, as well as the two {\em Chandra} results within 0.5--10 keV, are consistent. Assuming that the PWN-free off-pulse components is well represented by a simple power-law model, the PWN-free pulsed fraction is then
\begin{equation}
PF=\frac{KE^{-\alpha-\beta log(\frac{E}{E_{\rm p}})}}{KE^{-\alpha-\beta log(\frac{E}{E_{\rm p}})}+K_{off-pulse}E^{-\Gamma}}.
\label{PWN-freePF}
\end{equation}
Here $K_{off-pulse}$ and $\Gamma$ are the normalization factor and the photon index of the pulsar's PWN-free off-pulse component, to be determined. $K$, $\alpha$ and $\beta$ are the same as in Equation \ref{eq:logpar} and Table \ref{ta:pulsed}. With these four PWN-free pulsed fraction results, we took their upper and lower 1$\sigma$ limits and estimated that the off-pulse component's photon index is between 1.26 and 1.96. This constraint is weak due to the large uncertainties. {\em NuSTAR} observations with longer exposure time could be useful for constraining the off-pulse spectrum better. 

\section{Summary}
\label{sec:sum}
We have presented {\em NuSTAR} hard X-ray observations of the young rotation-powered radio pulsar PSR B1509$-$58 in the supernova remnant MSH 15$-$5{\it 2}. We have confirmed the curvature in the hard X-ray band reported by \cite{2001A&A...375..397C}, and have showed that the logpar model is statistically a better representation of our spectrum in both the {\em NuSTAR} energy band from 3 to 79 keV, and a broadband from 3 keV to 500 MeV. The logpar model predicts that the broadband energy spectrum peaks at $2.6\pm0.8$ MeV, $\sim$9 times more precisely determined than before, and consistent with the recent result of \cite{2015MNRAS.449.3827K}. We have also fitted the logpar model and the power-law model to the phase-resolved spectra in 26 phase intervals, but found that statistically the logpar model was not significantly superior to the power-law model in any of these phase intervals. In addition, we have measured the pulsed fraction of PSR B1509$-$58 in the hard X-ray energy band for the first time. The results imply that the PWN contributes $0.23\pm0.06$ of the total number of {\it NuSTAR} photons, while the pulsar's PWN-free off-pulse component contributes $0.09\pm0.07$ of the total number of photons. We have also estimated that the off-pulse component's photon index is between 1.26 and 1.96. To improve the constraint, {\em NuSTAR} observations with longer exposure time could be helpful. Finally, these new {\em NuSTAR} observations support a model (\citealt{0004-637X-764-1-51}) in which the X-rays originate from secondary pairs producing synchrotron radiation. The model describes how outgoing GeV emission goes undetected because of the Earth viewing angle. This helps to understand PSR B1509$-58$'s lack of GeV emission, and provides a possible answer to the long-lasting question about the source's atypical spectrum.

\medskip
This work was supported under NASA Contract No. NNG08FD60C, and made use of data from the {\em NuSTAR} mission, a project led by the California Institute of Technology, managed by the Jet Propulsion Laboratory, and funded by the National Aeronautics and Space Administration. We thank the {\em NuSTAR} Operations, Software and Calibration teams for support with the execution and analysis of these observations. This research has made use of the {\em NuSTAR} Data Analysis Software (NuSTARDAS) jointly developed by the ASI Science Data Center (ASDC, Italy) and the California Institute of Technology (USA). H.A. acknowledges supports provided by the NASA sponsored Fermi Contract NAS5-00147 and by Kavli Institute for Particle Astrophysics and Cosmology. V.M.K. acknowledges support from an NSERC Discovery Grant and Accelerator Supplement, the FQRNT Centre de Recherche Astrophysique du Quebec, an R. Howard Webster Foundation Fellowship from the Canadian Institute for Advanced Research (CIFAR), the Canada Research Chairs Program and the Lorne Trottier Chair in Astrophysics and Cosmology.

\bibliography{1509bib}
\bibliographystyle{apj}



\end{document}